\documentclass{article}

\usepackage{arxiv}

\usepackage[utf8]{inputenc} 
\usepackage[T1]{fontenc}    
\usepackage{hyperref}       
\usepackage{url}            
\usepackage{booktabs}       
\usepackage{amsfonts}       
\usepackage{nicefrac}       
\usepackage{microtype}      
\usepackage{lipsum}
\usepackage{graphicx}
\usepackage{comment}
\usepackage{float}
\usepackage{rotating}
\graphicspath{ {./images/} }

\title{Longitudinal Analysis of Heart Rate and Physical Activity Collected from Smartwatches}

\author{
 Fatemeh Karimi \\
  Department of Mathematical Sciences\\
  Sharif University of Technology\\
  Tehran, Iran \\
  \texttt{fatemeh.karimii@sharif.edu} \\
   \And
 Zohre Amoozgar \\
  Department of Radiation oncology\\
  Harvard Medical School\\
  Massachusetts, United States \\
  \texttt{zamoozgar@mgh.harvard.edu} \\
  \And
 Reza Reiazi \\
  Department of Radiation Physics\\
  University of Texas MD Anderson Cancer Center\\
  Texas, United States \\
  \texttt{rreiazi@mdanderson.org} \\
   \And
 Mehdi Hosseinzadeh \\
  Iran University of Medical Sciences\\
  Health Management and Economics Research Center\\
  Tehran, Iran \\
  \texttt{hosseinzadeh.m@iums.ac.ir} \\
  \And
 Reza Rawassizadeh  \\
  Department of Computer Science\\
  Boston University\\
  Massachusetts, United States \\
  \texttt{rezar@bu.edu } \\
}

\begin{document}
\maketitle

\begin{abstract}
Smartwatches (SWs) can continuously and autonomously monitor vital signs, including heart rates and physical activities involving wrist movement. The monitoring capability of SWs has several key health benefits arising from their role in preventive and diagnostic medicine. Current research, however, has not explored many of these opportunities, including longitudinal studies. In our work, we gathered longitudinal data points, e.g., heart rate and physical activity, from various brands of SWs worn by 1,014 users. Our analysis shows three common heart rate patterns during sleep but two common patterns during the day. We find that heart rate and physical activities are higher in summer and the first month of the new year compared to other months. Moreover, physical activities are reduced on weekends compared with weekdays. Interestingly, the highest peak of physical activity is during the evening.

\end{abstract}


\section{Introduction}
Digital medicine is gaining popularity and utility through the advent of wearable technologies as these devices are socially accepted throughout the globe, e.g., SWs and fitness trackers. These devices are becoming a highly ubiquitous type of technology, and have revolutionized the use of Holter devices. Wearable technologies are used to collect physiological parameters, such as heart rate, fitness activities, Galvanic Skin Resistance (GSR), and temperature, which are also helpful for monitoring users’ vital signs \cite{reeder2016health}. They can be used long-term, even by healthy individuals, constantly sensing, gathering, and reflecting on physiological data. 

Interestingly, smart everyday wearable technologies are the most popular devices as ranked by the consumers \cite{gan202111,majumder2017wearable,lu2020wearable}. SWs and fitness trackers (the most popular wearable devices) have constant contact with the skin \cite{rawassizadeh2014wearables}. This feature makes them valuable tools for tracking and improving users' health. The positive impact of SW is realized by assessing users' health through collecting biological, environmental, and behavioral information and quick access to them \cite{wang2021advances}.

Covid-19 pandemic \cite{channa2021rise} that limited one-one contact with healthcare professionals boosted the popularity of SWs and fitness trackers for health monitoring increased among end-users. In parallel, the cost of manufacturing SWs has dropped tremendously, increasing global access to these multi-functional devices \cite{visuri2017quantifying}. With a smartwatch, users can measure heart rate with their Photo-Plethysmo-Graphy (PPG) sensor and check their heart rate without expert monitoring or carrying extra devices \cite{rawassizadeh2019manifestation}. Additionally, accelerometers and gyroscope sensors function as a pedometer and count the number of steps to compute the consumed calories, allowing individual users to gain awareness of their dietary habits and the consequent physical health.

Altogether, more than 350,000 mobile health (mHealth) applications on the market \cite{murray2022digital}. A large group of existing mHealth applications focuses on automatically tracking the physical activities of users, which are known as fitness trackers. However, healthcare costs are increasing worldwide, especially in the U.S. \cite{poisal2022national}, which has a very high smartwatch penetration rate and is the largest consumer market for the wearable. This issue implies that there are still unexplored potentials in mHealth technologies. 

We believe long-term studies on users' heart rate and physical activity can tap into new behavioral patterns to mitigate health-related risks.

Our study aims to identify users' new behavioral patterns and behavioral dynamics of physical activity along with heart rate changes. Different stakeholders in this field could benefit from our results, including general health practitioners, application designers, device producers, and smartwatch end-users.

In this research, we used SWs with a pedometer to measure physical activity and PPG to measure heart rate in many users. We studied the dataset containing about 660,000 physical activity data points for 1,014 users and selected 322 users who chose to share their continuous heart rate variability data and 363 users who shared their physical activities; in total, we analyzed 685 users' data.

In particular, our contributions to the field of digital medicine are listed as follows.

\begin{itemize}
    \item We study heart rate variability during sleep time and the day to identify common patterns in heart rate variability at different times.
    \item We study different temporal patterns of physical activities, including changes in physical activities during the time of day, weekdays versus weekends, and different seasons.
    \item we experiment with different clustering and grouping methods and identify the most accurate clustering algorithm (Rand index and Purity) to group heart rate variability. 
\end{itemize}

\section{Related Works}
In this section, we highlight recent advancements in research that focus on leveraging SWs or fitness trackers for remote health monitoring and promoting well-being \cite{0m}. Since our work focuses on large-scale analysis of smartwatch data for health purposes, we also benefit from previous works that analyze smartwatch data. 

The first section focuses on smartphone and smartwatch approaches to analyze longitudinal health data. The second section highlights methodologies to analyze heart rate data collected from SWs, and the third section lists modalities that analyze physical activity data from SWs. We divided our related works into three sections based on the contributions of our research.

Given that other studies, \cite{jeong2017smartwatch,visuri2017quantifying,homayounfar2020understanding} relevant to longitudinal data of SWs are not collecting health data (heart rate and physical activity in our case); we did not include them in the discussion of our related work.    

\subsection{Longitudinal health studies on smartphones and SWs}

Mobile Health (mHealth) applications currently use battery-powered devices such as smartphones for personalized health monitoring \cite{5m,7m}. 
  
Many researchers investigated the impact of smartphones on health and the utility of smartphone features for wellness and wellbeing \cite{ahmad2016health,lane2011bewell,horwood2019problematic,cho2015roles,bakolis2018urban,mckay2019using}. There are apps focused on a single use case, such as managing relapses after quitting smoking \cite{1m}, tracking and managing diabetes \cite{2m}, such as monitoring nutrition, dietary behaviors, and food consumption \cite{3m}, sleep disorders \cite{ye2021fenet}, psychological distress \cite{cammisuli2021improving}, and etc. Sensors and characteristics of existing smartphones enable developers to build healthcare apps and allow caregivers to have continuous access to users' personal information, such as allergies, that can be checked daily by the smartphone application. Therefore, there is a social shift in the use of smartphones. Since the smartphone’s computing power supports other apps\cite{12m}, these devices cannot provide sufficient power of computing for sensors that require a constant connection to the skin, such as heart rate and galvanic skin resistance (GSR) and body temperatures \cite{13m}. However, wearable devices such as SWs or fitness trackers bridge this gap and enable continuous sensing data collection. This capability of SWs and fitness trackers makes them an ideal device for the continued collection of heart rate variability data and physical activity. These two sensors have many other practical applications in healthcare \cite{14m}. The exponential use of SWs and fitness trackers in healthcare \cite{jensen2021esc} and its rapid growth is other evidence of the efficiency of such wearable technology in improving the population's health. In general remote monitoring enables non-invasive monitoring of various types of physiological data and activities, such as stress\cite{4m}, sleep quality \cite{19m}, or physical activities and fitness status of users \cite{5m}.

\subsection{Analysis of heart rate }

Jovan on \cite{jovanov2015preliminary} work is among the early research work that leverages physiological monitoring performance of existing SWs to analyze users' health. While valuable, these findings have limitations due to the shortcomings of SWs at the time. The author employed SWs for longitudinal and continuous health monitoring for four months. Later there were other works \cite{gal2018effect,romeo2019can,rohani2018correlations} that explored SWs for health purposes. On the broader scale some \cite{tison2018passive,hicks2019best} analyze datasets of heart rate variability and some analysis physical activity \cite{holko2022wearable,tison2020worldwide,barbosa2018human}. However, they were analyzed independently, and no study has been done to identify  the correlation between the two approaches.

Identifying cardiac arrhythmia enables caregivers to recognize pulse irregularity and variability. An example is the Apple Heart \cite{perez2019large} that studies atrial fibrillation and applies an algorithm that uses pulse rate data to identify and evaluate "atrial fibrillation" in a large group of Apple Watch users. Another example is the apps that investigate the smartwatch for monitoring fluctuations in Parkinson's disease \cite{powers2021smartwatch,sigcha2021automatic,dubey2015echowear,sharma2014spark}. Using the device has proven to be efficacious in patient-clinician communication, allowing for monitoring motor symptoms in Parkinson's disease. Another group of studies \cite{patel2012review,chae2020development,thorpe2019adapting} focused on rehabilitating elderly users or individuals with chronic conditions that may lack the requisite ability to monitor health. Altogether, wearable devices maintain access to patient data enabling better monitoring and consequent medical interventions.

\subsection{Analysis of physical activity }

Physical activity monitoring using SWs \cite{p2,p3} can overcome the limitations of fitness trackers and incorporate individual feedback. Using \cite{althoff2017large} smartphone data analysis for the distribution of physical activity demonstrated a correlation between exercise, obesity prevalence, and geographical location. This report highlighted the significance of smartwatch use in improving physical activity. Another study \cite{p1} combined data from SWs and social media sharing to help users lose weight and effectively increase their motivation to perform physical activity. 

While these results are encouraging but they share a significant limitation. Despite the diversity of SWs in the market, previous studies relied on data collected with uniform devices. Analysis of specific brands does not account for heterogeneity while biasing toward device limitations and configurations. In our work, we used a wide range of smartwatch datasets, including heart rate, and physical activity, for a long time (almost five years).

\section{Method}
In this section, first, we describe the dataset we use for our analysis. Then, we describe our approaches to analyzing heart rate and physical activities as separate modalities. 

\subsection{Dataset}
We analyze a unique dataset collected using an Android SW application \footnote{https://play.google.com/store/apps/details?id=com.insight.insight}. The data collection process is IRB approved, and the privacy of a user who has agreed to participate and share their data is preserved. Besides, to respect users’ privacy we do not collect any demographic information, which is a limitation of our work. This study complies with the Guidelines for Accurate and Transparent Health Estimates Reporting (GATHER) \cite{stevens2016guidelines}. Our dataset includes eight different sensors \cite{rawassizadeh2015energy}. Still, we benefit from two of them in this study, “activity” that The Google Fit (https://developers.google.com/fit) API was used to gather this data and “heart rate” that was performed using native libraries of the Wear OS platform. More detail about the characteristics of each sensor with implementation and data collection principles are accessible in another work \cite{rawassizadeh2015energy}. 

If users consent to share their data, when a WiFi connection is available on the smartphone, data is uploaded to the smartphone from the SW, via Bluetooth, every three hours. Then the compressed data is uploaded to the privacy-protected cloud (AWS S3 instance). To retain SW storage data, data only retains on the device for ten days and afterward discarded automatically. The bi-modal data storage  mechanism is advantageous as reduces the uploading of large data packets to the network. \cite{homayounfar2020understanding}.

We benefited from covering 1,014 users worldwide distribution of the participants is represented in \ref{tab:region}. Users can choose to share or not share what types of information. This results in having a limited number of users sharing both their heart rate and physical activities, in total, we have 314 users who share both information, 322 users share only physical activities and 363 users share only heart rate variability. 

The default option in monitoring is to log  heart rate once every 30 minutes. Such a data recording results in 8,847 recordings in a week. Among these reports 3,328 records belong to the morning (between 6:00 a.m. and 12:00 p.m.); 3,314 recordings to the evening (between 12:00 p.m. and 6 p.m.), and 2,357 recordings at night (between 6 p.m. and 12 p.m) 3,702 on weekends, and 5,175 on the weekdays. Users can also configure how often the device records their heart rate (e.g.,10 minutes, 30 minutes, or one hour).

\begin{table}
\begin{center}
\caption{The geographical distribution of users by longitude and region. As seen from here, the number of users in the southern hemisphere is insignificant. Therefore, winter and summer weather is not contracting each other for different users.}\label{tab:region}
\begin{tabular}{|l|l|l|}
\hline
\multicolumn{1}{|c||}{Region} & \multicolumn{1}{|c||}{Users(in Percentage)}\\
\hline
\multicolumn{1}{|c||}{Europe} & \multicolumn{1}{|c||}{47.92} \\
\multicolumn{1}{|c||}{North America} & \multicolumn{1}{|c||}{41.11}\\
\multicolumn{1}{|c||}{Asia} & \multicolumn{1}{|c||}{10.21}\\
\multicolumn{1}{|c||}{Australia\& Ocean}  & \multicolumn{1}{|c||}{0.63}  \\
\multicolumn{1}{|c||}{South America} & \multicolumn{1}{|c||}{0.13}\\
\multicolumn{1}{|c||}{North Africa} & \multicolumn{1}{|c||}{0.601}\\
\hline
\end{tabular}
\end{center}
\end{table}

\subsection{Grouping Common Patterns}

The objective of this study is to identify common patterns of heart rate changes, and temporal dynamics of human activities using smartwatch sensors. The smartwatch sensor data comes from the smartwatch's pedometer (accelerometer, gyroscope), and heart rate (PPG) sensors. The heart rate data is considered as a daily time series from the beginning of the day (00:00) to the end of the day (23:59).
To identify these patterns, we apply time-series clustering to find common patterns in heart rate. 

We investigate various methods of clustering, including partition-based clustering (k-mean \cite{day1984efficient} and k-shape \cite{paparrizos2015k} and kernel k-means \cite{dhillon2004kernel}), density-based clustering (DBSCAN\cite{ester1996density} and OPTICS \cite{ankerst1999optics}), and agglomerative hierarchical clustering, Ward \cite{murtagh2011ward}. Besides, we employ Self-Organized-Maps (SOM) for projecting data into lower dimensions and then applying SOM clustering, based on distances between data points. 

\subsubsection{Rational of Clustering Algorithm Selection}
Since the clusters may have a non-elliptical shape, we used DBSCAN and OPTICS density-based clustering methods\cite{bar2007non}. DBSCAN and OPTICS account for the adjacent areas of high appearance density that are separated from other clusters by low density and also noise or outliers. 

Since the clusters may have a non-elliptical shape, we used DBSCAN and OPTICS density-based clustering methods \cite{bar2007non}. DBSCAN and OPTICS account for the adjacent areas of high appearance density that are separated from other clusters by low density and also noise or outliers. 

Agglomerative hierarchical clustering \cite{macqueen1967classification} is a type of hierarchical clustering algorithm that builds a cluster structure using the bottom-up approach and each cluster can have several sub-clusters. Since we suspect some behavioral patterns could have hierarchical characteristics, we also used agglomerative for clustering. 

To implement partition-based clustering we use k-mean and k-shape algorithms. The k-means clustering algorithm is one of the most popular unsupervised clustering algorithms. Another  partition-based clustering is the k-shape algorithm. It measures distances based on the shape defined\cite{paparrizos2015k} during the time series clustering. Since it uses cross-correlation it is not sensitive to shift and scale. Another algorithm in this category is Kernel k-means\cite{dhillon2004kernel} is considered a version of the k-means algorithm that transforms the input space to a high-dimensional feature space where the nonlinear separators apply in this space.

The Self-Organizing-Map uses a one-layer neural network to extract features from time series\cite{vesanto2000clustering}, and project the result into a lower-dimensional space. Due to its superior accuracy, we use it for clustering heart rate time series. We train a SOM neural network with heart rate variability data. 

\subsection{Statistical Analysis}
We examined the correlation between heart rate and physical activity, by using Pearson, Kendal Tau, and Spearman correlation metrics. We do not get a high correlation score. This is due to the fact that heart rate changes depends on many factors, such as age \cite{van1993heart}, cardiovascular disease \cite{fox2007resting}, high cholesterol \cite{thayer2010relationship}, diabetes \cite{cheng2003heart}, emotional state (e.g. stress) \cite{anttonen2005emotions}. Many contributing factors lead to complex heart rate patterns that are not necessarily correlated with fitness activities. Therefore, fitness and physical activity level are one of them and can not establishes a direct correlation with the data.
\begin{table}
\begin{center}
\caption{Silhouette scores for k-means, k-shape, and kernel k-means to identify the optimal number of clusters.} \label{tab:si}
\begin{tabular}{|l|l|l|l|}
\toprule
\multicolumn{4}{c}{\textbf{Silhouette Index}} \\ \midrule
\hline
\multicolumn{1}{|c||}{num. of clusters} & \multicolumn{1}{|c||}{k-means} & \multicolumn{1}{|c||}{k-shape} & \multicolumn{1}{|c||}{kernel k-means} \\
\hline
\multicolumn{1}{|c||}{2} & \multicolumn{1}{|c||}{0.136} & \multicolumn{1}{|c||}{0.11} & \multicolumn{1}{|c||}{0.13} \\
\multicolumn{1}{|c||}{3} & \multicolumn{1}{|c||}{0.056} & \multicolumn{1}{|c||}{0.11} & \multicolumn{1}{|c||}{0.14}\\
\multicolumn{1}{|c||}{4} & \multicolumn{1}{|c||}{0.058} & \multicolumn{1}{|c||}{0.055} & \multicolumn{1}{|c||}{0.14}\\
\multicolumn{1}{|c||}{5} & \multicolumn{1}{|c||}{0.394} & \multicolumn{1}{|c||}{0,052} & \multicolumn{1}{|c||}{0.12}\\
\multicolumn{1}{|c||}{6} & \multicolumn{1}{|c||}{0.051} & \multicolumn{1}{|c||}{0.012} & \multicolumn{1}{|c||}{0.14}\\
\multicolumn{1}{|c||}{7} & \multicolumn{1}{|c||}{0.040} & \multicolumn{1}{|c||}{0.028} & \multicolumn{1}{|c||}{0.12}\\
\multicolumn{1}{|c||}{8} & \multicolumn{1}{|c||}{0.064} & \multicolumn{1}{|c||}{0.014} & \multicolumn{1}{|c||}{0.11}\\
\multicolumn{1}{|c||}{9} & \multicolumn{1}{|c||}{0.044} & \multicolumn{1}{|c||}{0.010} & \multicolumn{1}{|c||}{0.10}\\
\multicolumn{1}{|c||}{10} &\multicolumn{1}{|c||}{0.027} & \multicolumn{1}{|c||}{-0.093}&\multicolumn{1}{|c||}{0.10} \\
\multicolumn{1}{|c||}{11} &\multicolumn{1}{|c||}{0.021} & \multicolumn{1}{|c||}{-0.045} &\multicolumn{1}{|c||}{0.096}\\
\multicolumn{1}{|c||}{12} &\multicolumn{1}{|c||}{0.019} & \multicolumn{1}{|c||}{-0.057} &\multicolumn{1}{|c||}{0.095}\\
\multicolumn{1}{|c||}{13} &\multicolumn{1}{|c||}{0.018} & \multicolumn{1}{|c||}{-0.022}& \multicolumn{1}{|c||}{0.095}\\
\multicolumn{1}{|c||}{14} &\multicolumn{1}{|c||}{0.022} & \multicolumn{1}{|c||}{-0.021}& \multicolumn{1}{|c||}{0.084} \\
\multicolumn{1}{|c||}{15} &\multicolumn{1}{|c||}{0.023} & \multicolumn{1}{|c||}{-0.091} & \multicolumn{1}{|c||}{0.088}\\
\multicolumn{1}{|c||}{16} &\multicolumn{1}{|c||}{0.005} & \multicolumn{1}{|c||}{-0.020} & \multicolumn{1}{|c||}{0.075}\\
\multicolumn{1}{|c||}{17} &\multicolumn{1}{|c||}{0.010} & \multicolumn{1}{|c||}{-0.0027} & \multicolumn{1}{|c||}{0.070}\\
\multicolumn{1}{|c||}{18} &\multicolumn{1}{|c||}{0.063} & \multicolumn{1}{|c||}{-0.091}& \multicolumn{1}{|c||}{0.084}\\
\multicolumn{1}{|c||}{19} &\multicolumn{1}{|c||}{0.0017} & \multicolumn{1}{|c||}{-0.0201} & \multicolumn{1}{|c||}{0.084}\\
\multicolumn{1}{|c||}{20} &\multicolumn{1}{|c||}{0.0018} & \multicolumn{1}{|c||}{-0.0027}& \multicolumn{1}{|c||}{0.082}\\
\hline
\end{tabular}
\end{center}
\end{table}

\section{Result}
All results reported in this section and a comparison between two or more a group of data have been verified with a statistical significance test. In other words, we achieve $p-value < 0.05$, by using KS-Test for all reported results and we do not report results that are not statistically significant.
We compare all of the time series clustering algorithms that we have listed in the method section and evaluate the quality of clusters to obtain the best result that leads to identifying behavioral patterns.

\begin{figure}[H]
  \centering
  \includegraphics[width=8 cm]{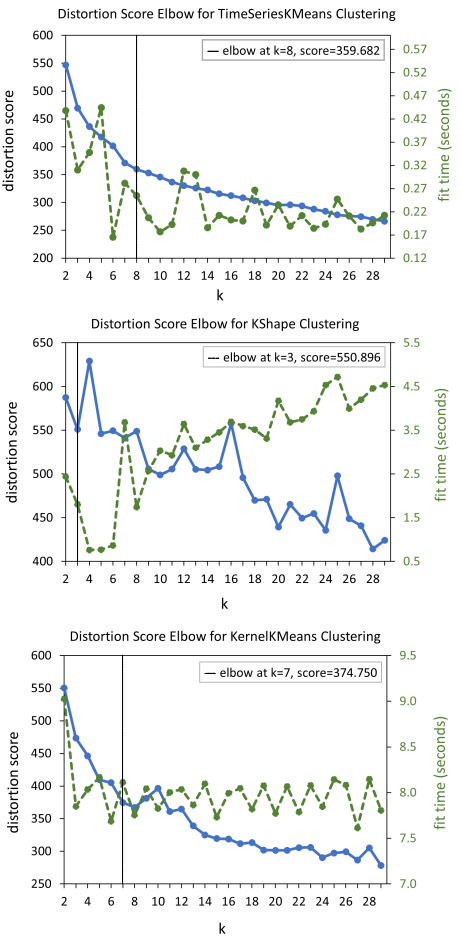}
  \caption{The elbow method, based on distortion score, for identifying the optimal number of clusters in three methods: k-means, k-shape, and kernel k-means.}\label{fig:eb1}
\end{figure}

We experiment with different clustering configurations and  evaluate the result. This enables us to select the best  clustering algorithm with the best parameter settings.

\subsection{Intrinsic Evaluation}
A fundamental stage for any clustering approach is to determine the number of clusters. First, we analyze partitioning methods using the Silhouette Index and  Elbow method, common approaches to calculating the intrinsic measurement. Afterward, we find the optimal node for SOM.

\subsubsection{Silhouette Index}
To identify the optimal number of clusters while using partitioning-based clustering, we use the Silhouette Coefficient \cite{struyf1997clustering} that provides a quantitative score using the mean intra-cluster distance and the mean nearest-cluster distance ranging from -1 to +1. In particular, it performs a similarity comparison between clusters that separate clusters from each other. Table \ref{tab:si} shows the average silhouette score for the k-means, kernel k-means, and k-shape algorithms.
\subsubsection{Elbow Method}
Another common method in cluster analysis is the elbow method \cite{thorndike1953belongs} which determines the optimal number of clusters into which the data may be clustered.
The method consists of plotting that considers variation as a relation between the number of clusters and a cost function. By observing the plot, we can pick the elbow of the curve as the optimal number of clusters. 

As shown in Figure \ref{fig:eb1} the elbow in k-means is eight, in kernel k-means is seven and in k-shape is three. Nevertheless, due to high fluctuations of k-shape clustering, we can not rely on the elbow method for this algorithm.

\subsubsection{Evaluating SOM}
To our knowledge, there is no ultimate solution exists for identifying an optimal number of clustering in SOM. There are promising studies \cite{tian2014anomaly,akccapynar2014investigating} that determine the number of clusters, but they are very dataset dependent. We experimented to evaluate the quality of SOM clustering and finally apply this equation: $num-clusters = \left \lceil \sqrt{\left \lceil \sqrt{num-users} \right \rceil} \right \rceil^{2}$, that this formula has the better result for our dataset and can get it direct from this link kaggle\footnote{\url{https://www.kaggle.com/code/izzettunc/introduction-to-time-series-clustering/notebook}} that analyzes time-series clustering.

\subsection{Extrinsic Evaluation}
In addition to intrinsic evaluation, we employ human subject to measure the quality of clustering and compares cluster labels with the expert annotated label. To conduct this experiment, two human experts annotate cluster content, with Cohen's kappa of 0.97, which is almost perfect agreement. 

\subsubsection{Purity}

A common approach to determine the quality of a clustering is cluster purity \cite{zhao2001criterion} which is a transparent validation measure. It compares the output of the clustering algorithm with the ground truth dataset, which is labeled manually by human subjects.

To conduct this evaluation a confusion matrix of size $2 \times 2$ which computes the similarity between two of its inputs is constructed. After that utilizing the Purity, a score  between 0 and 1 is provided, one indicates a high purity score (good clustering result), and zero indicates a low purity score (weak clustering result). Table \ref{tab:ex} presents results, SOM achieves the highest score and density-based clustering methods (DBSCAN and OPTICS) get the lowest score among other clustering methods.
\subsubsection{Rand Index}
Another common method for qualitative evaluation is the Rand Index \cite{rand1971objective}, which also relies on the ground truth dataset.  Rand Index is a measure that evaluates the match between clustering and ground truth. It provides a score between zero and one, when the score is close to one, is more consistent with ground truth, while a score close to zero is a sign of randomness. 

Density-based clustering methods receive the lowest Rand Index score and the SOM achieves the highest score (similar to Purity measurement). Also, partition-based methods have high Rand Index values close to each other.

\begin{figure}[H]
  \centering
  \includegraphics[width=\linewidth]{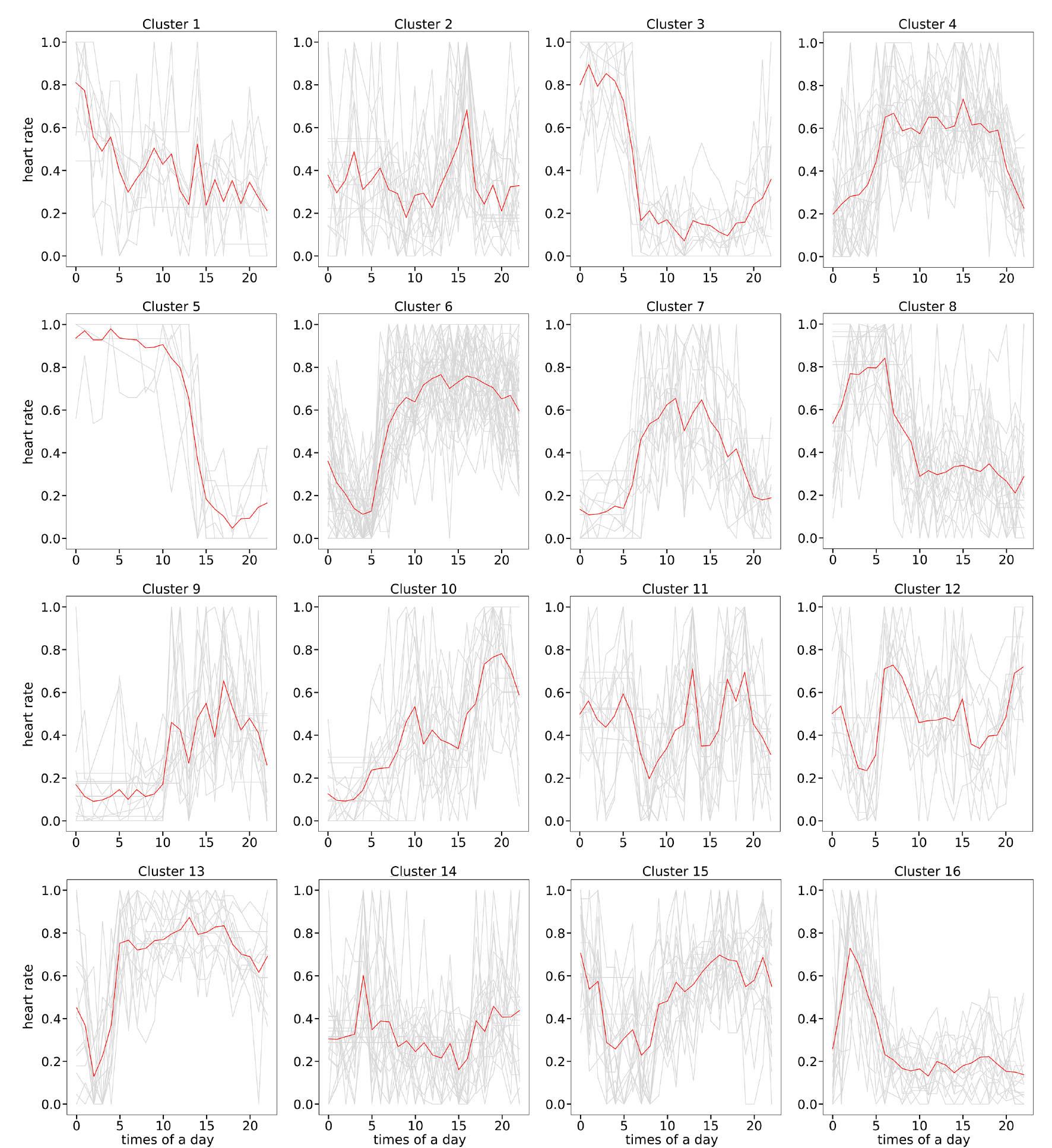}
  \caption{Results of heart rate clustering by using SOM algorithm.}
  \label{fig:full}
\end{figure}

Based on the result acquired from qualitative evaluations (Purity and Rand Index) and quantitative ones (Silhouette Index and Elbow Method), we select the SOM as the clustering approach and apply it to the dataset after 10000 iterations. The result achieved strongly separated clusters of data, and it divides the dataset into 16 result groups, see Figure \ref{fig:full}.

In this Figure, each cluster is summarized with a red curve as an average of all its time series in the cluster. The horizontal axis shows the time from "00:00" in the morning to "23:59" at night. The vertical axis shows the heart rate, mapped from the interval (31-245) to the interval (0-1).

\begin{table}
\begin{center}
\caption{Results of extrinsic evaluation methods (Rand index and Purity).}\label{tab:ex}
\begin{tabular}{|l|l|l|}
\hline
\multicolumn{1}{|c||}{Algorithm} & \multicolumn{1}{|c||}{Rand Index} & \multicolumn{1}{|c||}{Purity}\\
\hline
\multicolumn{1}{|c||}{SOM} & \multicolumn{1}{|c||}{0.903} & \multicolumn{1}{|c||}{0.24} \\
\multicolumn{1}{|c||}{Kmeans}& \multicolumn{1}{|c||}{0.894} & \multicolumn{1}{|c||}{0.22}\\
\multicolumn{1}{|c||}{Kshape} & \multicolumn{1}{|c||}{0.879} &  \multicolumn{1}{|c||}{0.20}\\
\multicolumn{1}{|c||}{kernelKmeans} & \multicolumn{1}{|c||}{0.885}& \multicolumn{1}{|c||}{0.22}\\
\multicolumn{1}{|c||}{OPTICS} & \multicolumn{1}{|c||}{0.451} & \multicolumn{1}{|c||}{0.14}\\
\multicolumn{1}{|c||}{BDSCAN} & \multicolumn{1}{|c||}{0.691} & \multicolumn{1}{|c||}{0.14}\\
\multicolumn{1}{|c||}{Agglomerative} & \multicolumn{1}{|c||}{0.832}& \multicolumn{1}{|c||}{0.16}\\
\hline
\end{tabular}
\end{center}
\end{table}

\section{Discussions and Findings}

\subsection{Heart rate Analysis}
Our  analysis shows that 62\% of users have a heart rate between 60 and 100, which means their average heart rate is normal, 21\% have above average upper bound or 100 (fast heart rate) and the rest of the users are below 60 or average lower bound (slow heart rate). 

Through using statistical analysis and time series clustering, We have analyzed the heart rate variation of users in the different time scopes. In particular, we investigate the heart rate at different times, (i) during the day, (ii) during a month, and (iii) months a year.

All findings we report are statistically significant  $( p-value < 0.05 )$, otherwise we mention them explicitly.
Our analysis reveals the average heart rate of users is 73 bpm at night (6 p.m. to 12 p.m.) and slower than  at other times, such as in the morning time ( 6 a.m. to 12 noon) and in the afternoon time (12 noon to 6 p.m.). In other words, the average heart rate of most users is decreasing during the day that heart rate in the morning is 80 bpm and in the afternoon is 77 bpm. 

\begin{figure}[H]
  \centering
  \includegraphics[width= 6 cm]{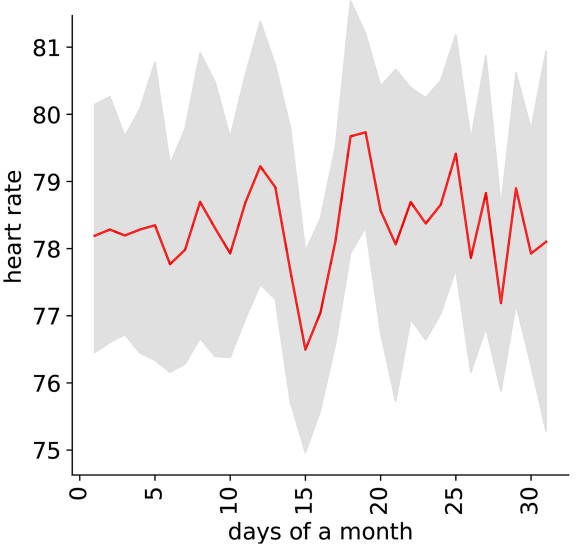}
  \caption{The average heart rate variability of all users, based on the day of a month.}
  \label{fig:ww1}  
\end{figure}

Analysis of users' heart rate during the days of a month shows the average heart rate in the middle of the month is at the lowest (Figure\ref{fig:ww1}). We did not identify any proper justification for this phenomenon and we leave it for future investigation. 

Our analysis reveals the average heart rate of users is 73 bpm at night (6 p.m. to 12 p.m.) and slower than  at other times, such as in the morning time ( 6 a.m. to 12 noon) and in the afternoon time (12 noon to 6 p.m.). In other words, the average heart rate of most users is decreasing during the day that heart rate in the morning is 80 bpm and in the afternoon is 77 bpm. 

Analysis of users' heart rate during the days of a month shows the average heart rate in the middle of the month is at the lowest (Figure\ref{fig:ww1}). We did not identify any proper justification for this phenomenon and we leave it for future investigation.

Studying heart rate variability patterns based on the month of the year shows that the average heart is highest in summer, and at the beginning of the year, which could be related to vacation times, when users have less work and more time for physical activities.  As it is shown in Figure \ref{fig:ss}, August has the highest heart rate, and March has the lowest one. The lowest value is in the fall (see Figure \ref{fig:ss}).

\begin{figure}[H]
  \centering
  \includegraphics[width= 9 cm]{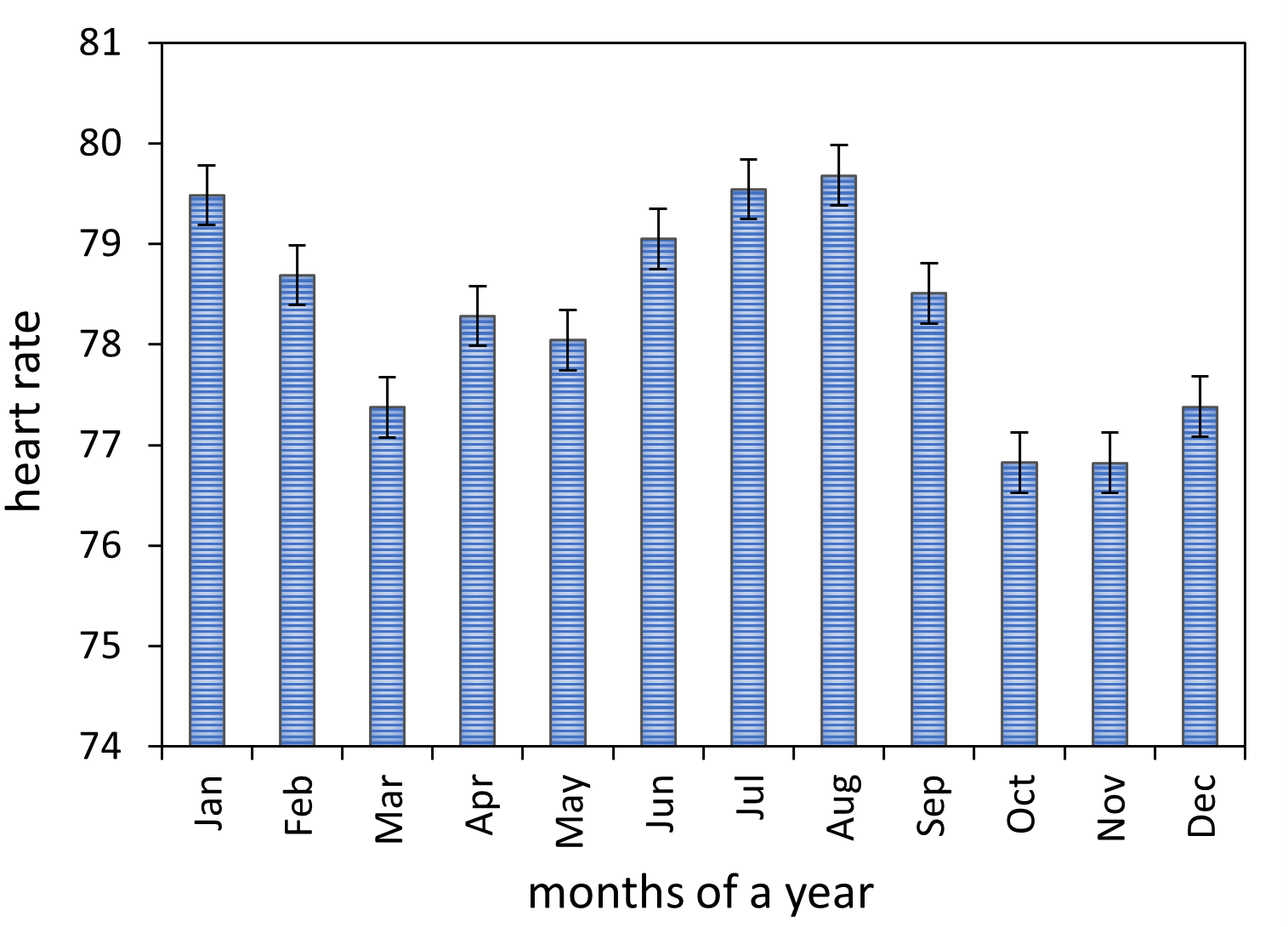}
  \caption{The average heart rate variability of all users, based on the  months of the year.}
   \label{fig:ss}
\end{figure}

\begin{figure}[H]
  \centering
  \includegraphics[width=9 cm]{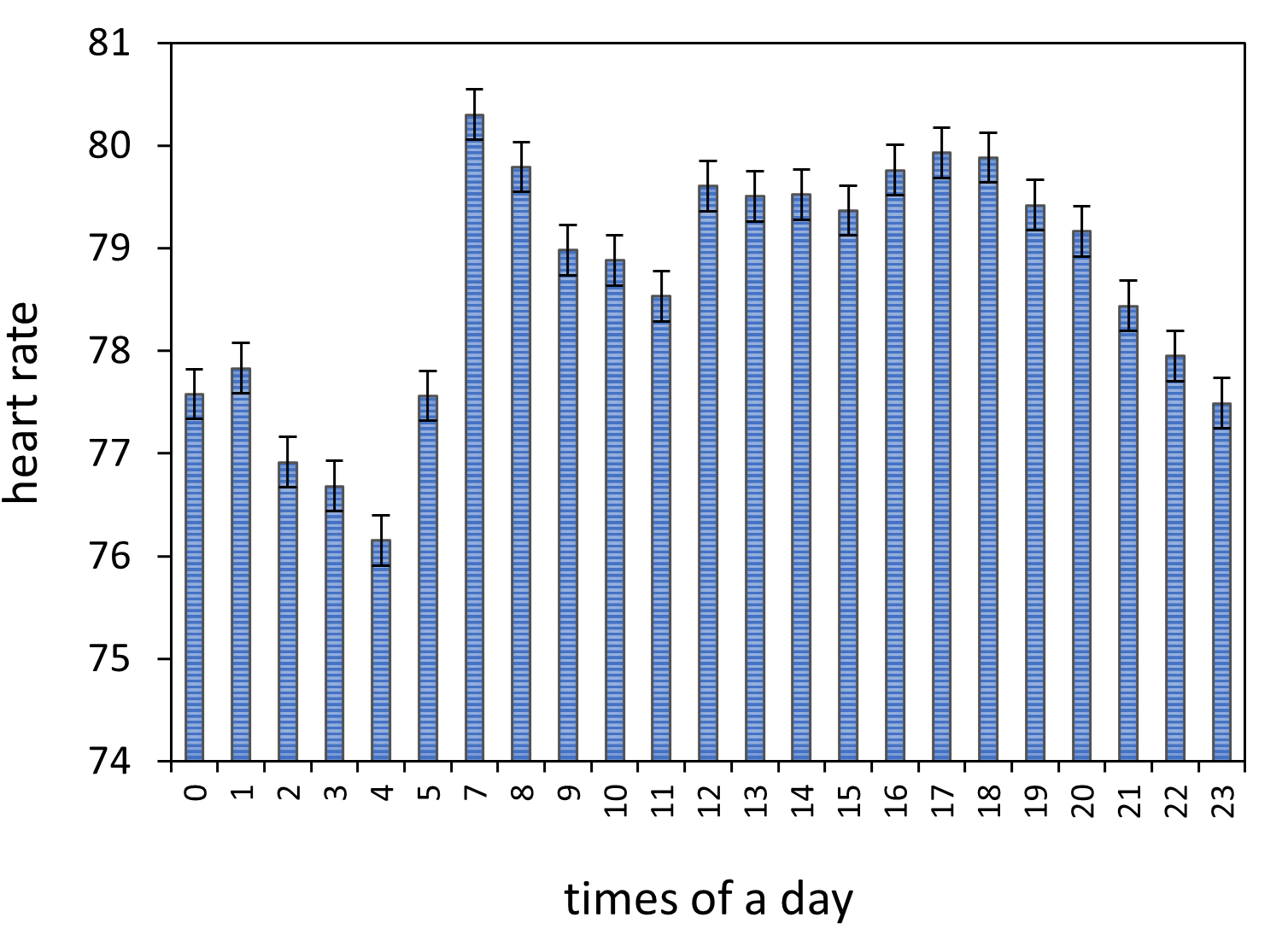}
  \caption{The average heart rate variability of all users, based on the time of the day.}
   \label{fig:ss}
\end{figure}

Our examination of the heart rate during the day shows that the heart rate decreases during sleep time and reaches its lowest value at around 4:00 a.m. and its highest value around 7:00 am in the morning.  Furthermore, the heart rate decreases during the day from 7 a.m. to 11 a.m. and is almost constant until 6 p.m. in the evening. Afterward, it decreases until the end of the day.

In the next sections, by examining the time series clustering, we investigate the reason for these peaks.
Next, we experiment with time series clustering of heart rate data at various times (sleep time, daytime, different seasons, weekends vs weekdays) and examine the findings in each part. 
The normal range of individual heart rate is between 60 and 100 which varies from person to person \cite{spodick1992operational}, but different factors such as physical activity \cite{strath2000evaluation}, hydration level \cite{carter2005influence}, and body temperature \cite{berggren1950heart} affect heart rate and our dataset is unable to capture all those details. Therefore, we describe our results without considering these factors.

\subsubsection{Heart Rate Patterns During the Sleep Time}
We study the heart rate variability, as a time series, during sleep time (from 12 a.m. to 8 a.m. which the timing is recommended by another work \cite{rawassizadeh2016scalable}), which is an important factor to check the heart health status and its functions \cite{cincin2015effect}. Furthermore, blood pressure is directly correlated with heart rate variability \cite{george1985sleep}, and thus investigating heart rate variability have several applications in users' health \cite{padwal2019accuracy}.

There are 29 users who are physically active during their sleep time and we do not include them in the study. We suspect that their job has night shifts, and thus they are not sleeping during the night time.  

\begin{figure}[H]
  \centering
  \includegraphics[width= 8 cm]{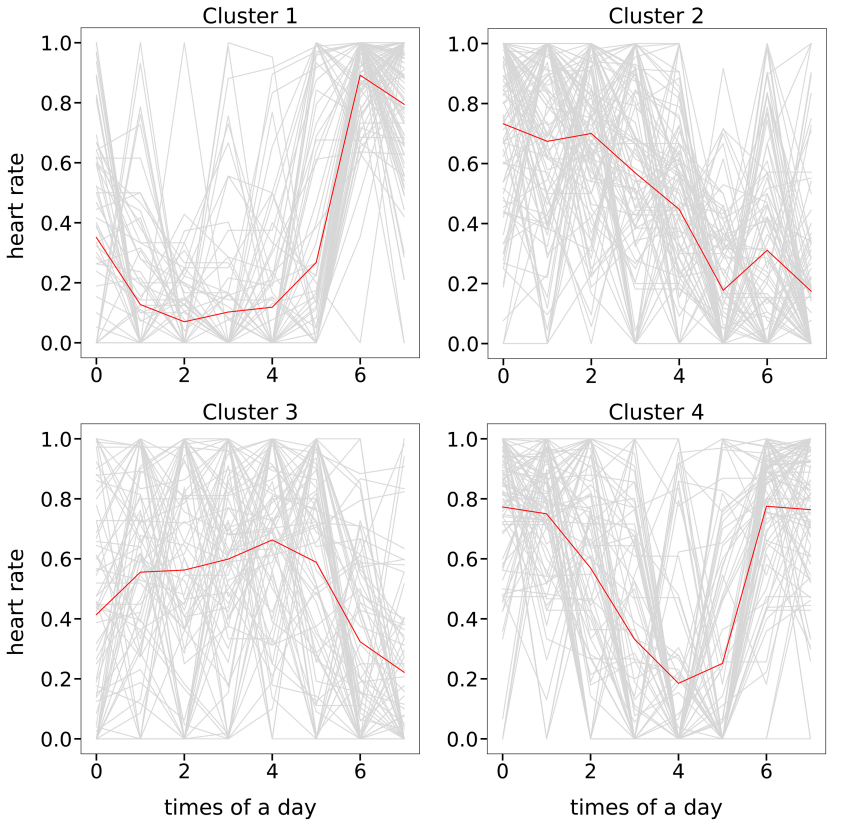}
  \caption{Results of the SOM clustering of heart rate during sleep time. Note that Cluster 2 and Cluster 3 present fairly similar patterns, and thus we summarize them in three patterns.}
  \label{fig:night}
\end{figure}

To identify patterns of heart rate variability during sleep time, we study the sloping trend of the heart rate curve,  the highest, and lowest point in the curve. As a result, we identify three significant heart rate patterns during sleep time; (i) \textbf{valley pattern}, (ii) \textbf{downward pattern}, and (iii) \textbf{peak pattern}, these patterns are visualized in Figure \ref{fig:night}. The heart rate patterns we have obtained via time series clustering from our dataset, are similar to findings that were measured with accurate Holter devices used for medical purposes \cite{stein2003simple,maier2014robust,stein2012heart}. This evidence could verify that even SWs that are not accurate can be used instead of expensive Holter devices used for health tracking, with a reasonable performance. Note that a user might have different patterns during different nights, but we report the most frequent pattern for each user.

\textbf{valley pattern:} indicates that the body organs relax in the early stages of sleep, and thus heart rate and blood pressure begin to decrease. The heart rate curve reaches the lowest heart rate in the middle of sleep when the melatonin levels peak \cite{griefahn2002validity}. The valley pattern is known as the ideal progression of heart rate during sleep \cite{smith2013influence}. When the body is synchronized with the patterns of the sun, the body temperature gradually decreases\cite{szymusiak2018body}. When waking up in the morning, the heart rate starts to increase due to the stress hormone cortisol\cite{hynynen2011incidence}, and at the end of sleep, the heart rate increases and peaks after waking up\cite{hynynen2009heart}. As a result, the valley pattern shows a high-quality sleep that the body has relaxed during the night \cite{smith2013influence} and the lower resting heart rate that we observe in this pattern is a symptom of healthy sleep\cite{fox2007resting}. 44\% of users have a valley pattern, which means that they have a normal sleep rhythm, which is an important part of a healthy lifestyle \cite{tanaka2004sleep}. 

\textbf{downward pattern:} This pattern shows a high heart rate at the beginning of sleep and reaches its lowest point just before waking up. This phenomenon indicates that the metabolism is visibly at work \cite{bannai2014association} and therefore the body feels weak \cite{nakamura1998increases}at the beginning of the day and blood pressure drops \cite{george1985sleep}. Late dinner time \cite{azami2019long}, and drinking alcohol before bed \cite{irwin2006association} are some of the important factors causing this sleep pattern. 
29.5\% users' data show a downward pattern in their sleep time.

\textbf{peak pattern:} This pattern shows an increase in heart rate just after falling asleep, and once reaching its peak, then it decreases. This pattern is known to cause some problems in physical or mental health \cite{azza2020stress,shinar2006autonomic}. A common  reason is related to late sleeping  and exhaustion during the day \cite{hynynen2006heart}. If the subject sleeps  on time and still this pattern persists, the heart rate may increase during sleep due to high stress or anxiety \cite{brosschot2007daily}, or a nightmare occurs during the sleep time \cite{perogamvros2019increased}.
26.3\% of users show a peak pattern.

\subsubsection{Heart Rate Patterns During the Day time}
\begin{figure}[H]
  \centering
  \includegraphics[width= 8 cm]{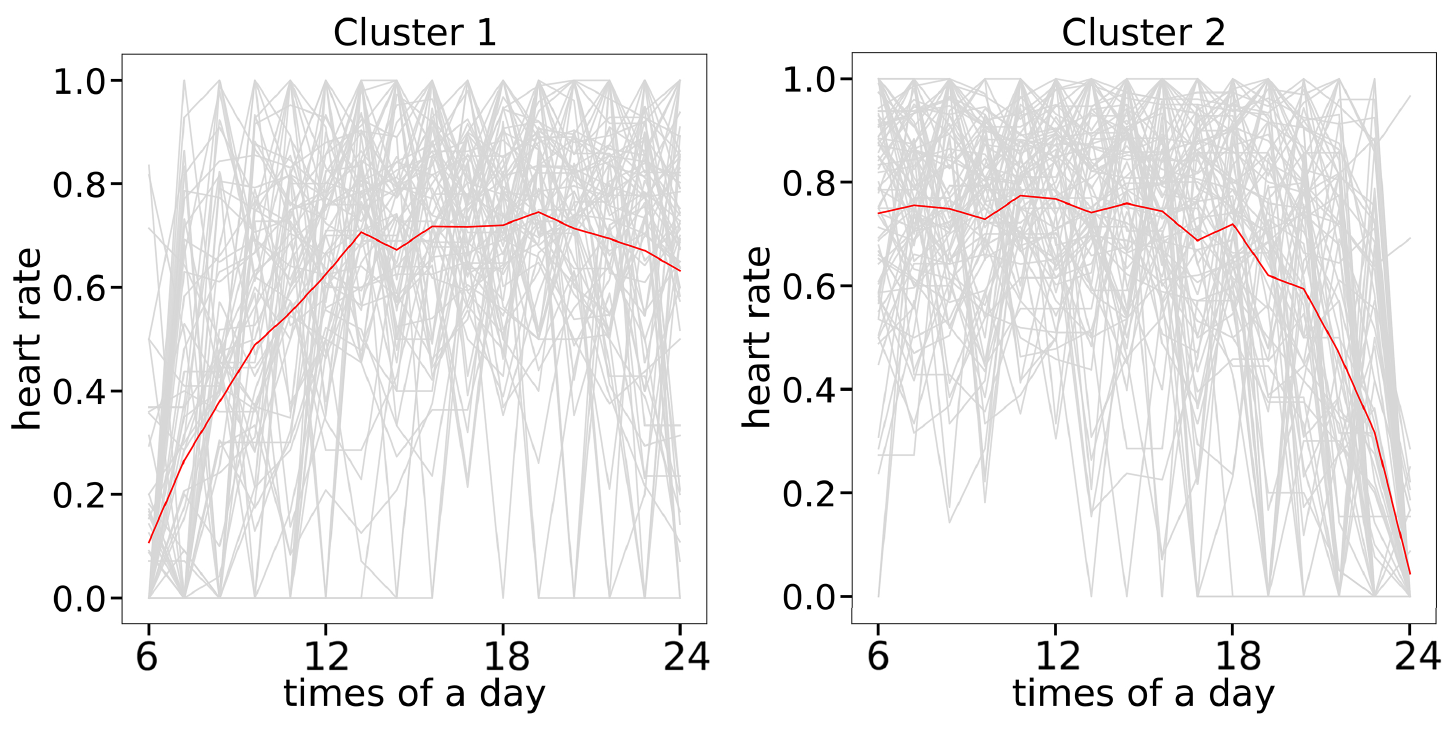}
  \caption{Results of heart rate clustering,  during the daytime, by using the SOM algorithm.} 
  \label{fig:day}
\end{figure}

The result of clustering revealed, two main patterns for heart rate variability during the day (6 a.m. to 12 a.m.), i.e., (i) \textbf{downward trend} and (ii) \textbf{upward trend}, These two patterns are presented in Figure \ref{fig:day}. The downward trend shows heart rate decreases at the end of the day when the body is relaxed during the rest period \cite{kang2017effect} and the quality of night sleep affects heart rate during the day.
In 78\% clusters, during the daytime, where the sleep pattern is the peak, the downward pattern occurs. These two sleep patterns are known as not of good quality \cite{lund2010sleep} and affect the body system during the day. 

In particular, as the heart rate drops, so does the blood pressure, the weakness of the body continues \cite{butler2004measuring} and causes a lack of energy \cite{wiklund2009influence} and low stamina \cite{udovcic2017hypothyroidism}. There are various reasons for reduced heart rate including  changes in cardiac-vagal tone, representing mechanical, physiological, and pathological alteration in the parasympathetic nervous system that regulates cardiac function. Another is hypothyroidism \cite{lin1992relationship}, in which the thyroid gland does not produce enough thyroid hormones \cite{cojic2017subclinical}.

\begin{figure}[H]
  \centering
  \includegraphics[width=\linewidth]{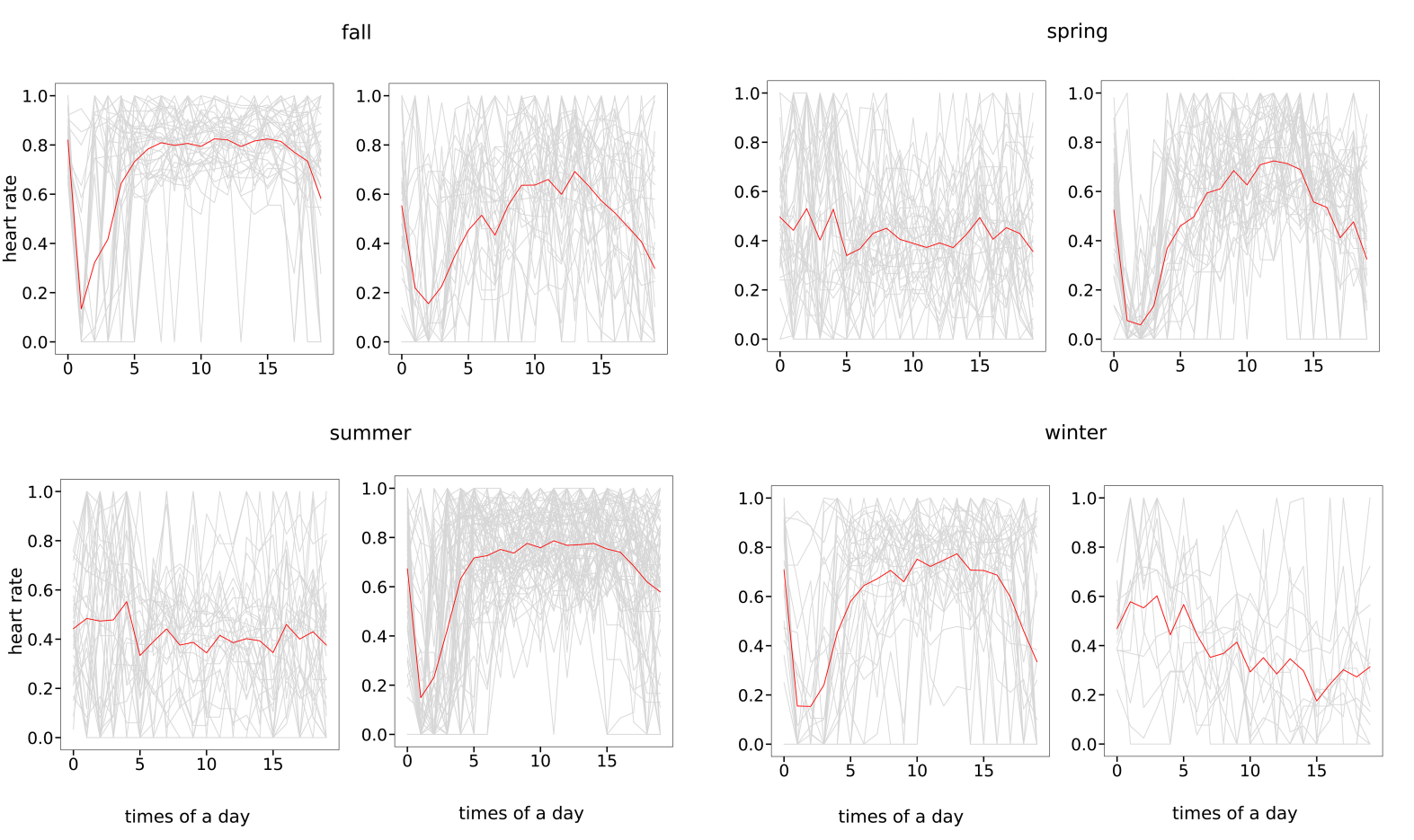}
  \caption{Results of SOM clustering based on seasons, i.e. common seasonal patterns.}
  \label{fig:nn}
\end{figure}
Hypothyroidism can affect the health of blood vessels \cite{klein2007thyroid}, which in turn can cause a slow heart rate \cite{buchheit2004effect}. The other potential reason is hypoxia \cite{klein2007thyroid} which happens when the body does not receive enough oxygen \cite{barrionuevo1999o2}.

41\% of users show a downward trend during the daytime. Given that we had no access to health records, and because the regulation of heart rate is multifactorial, we only provided examples of potential causes.

The upward trend of heart rate can originate from the increase in physical activity(\ref{fig:a0}), exercises significantly increase heart rate\cite{bernardi1996physical}. Moreover, when there is no physical activity, there are other reasons such as warm temperature, \cite{farrell1997effects,lyman1958oxygen}, smoking and tobacco consumption \cite{edwards1987effects,el2015health} that make the heart beat faster. Stress is another major factor contributing to the increase of heart rate, the higher the stress level, the faster the heart beats \cite{engert2016boosting}, 37\% of users show an upward trend during the daytime.

\subsubsection{Seasonal Patterns}
The sleep time patterns that we examined in the previous section occur in different seasons according to the conditions of each season.

 We study the heart rate during sleep time (from 12 a.m. to 8 a.m.) in four seasons of the year. Our analysis revealed, \emph{during the fall, the valley pattern. which is a sign of a healthy sleep routine is at its highest}, $p-value < 0.5$. A reason we speculate is the mild climate that regulates the heart rate.
 
 In winter and summer, the heart rate variability during sleep time is not significantly different and includes two main patterns; valley (62\%) and peak (31\%) happens. In spring in addition to the valley pattern (51\%), we observe the downward pattern (47\%) as well, and the heart rate decreases during sleep time. 

\subsubsection{Weekend versus Weekday patterns}

We examine the heart rate on weekdays and weekends. Given that most of our users were localized in countries with Saturday and/or Sunday as weekends, we considered Sunday as the unifying weekend. 

\begin{figure}[H]
  \centering
  \includegraphics[width=\linewidth]{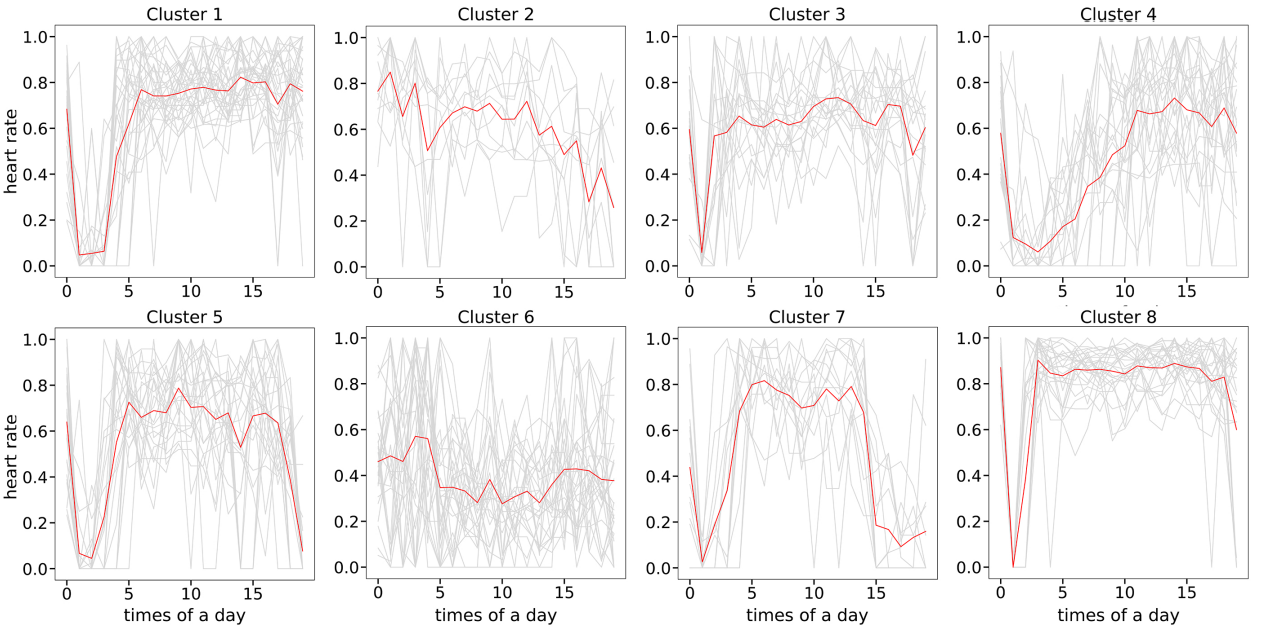}
  \caption{Results of heart rate clustering for weekends, by using SOM algorithm.}
  \label{fig:wend}
\end{figure}

First, we found that heart rate on Sunday nights includes three patterns(figure \ref{fig:wend}); the valley pattern that 63\% users have the valley sleep behavior and which is a sign of proper sleep, the peak pattern (cluster 6) that occurs when users sleep late, 14\%, and the last pattern is downward (cluster 2), i.e. 11\%. It might be either exhaustion during the week that affects sleep on weekends \cite{noland2009adolescents}, or users' nightlife, as it is shown in Figure \ref{fig:wend}.

\begin{figure}[H]
  \centering
  \includegraphics[width=\linewidth]{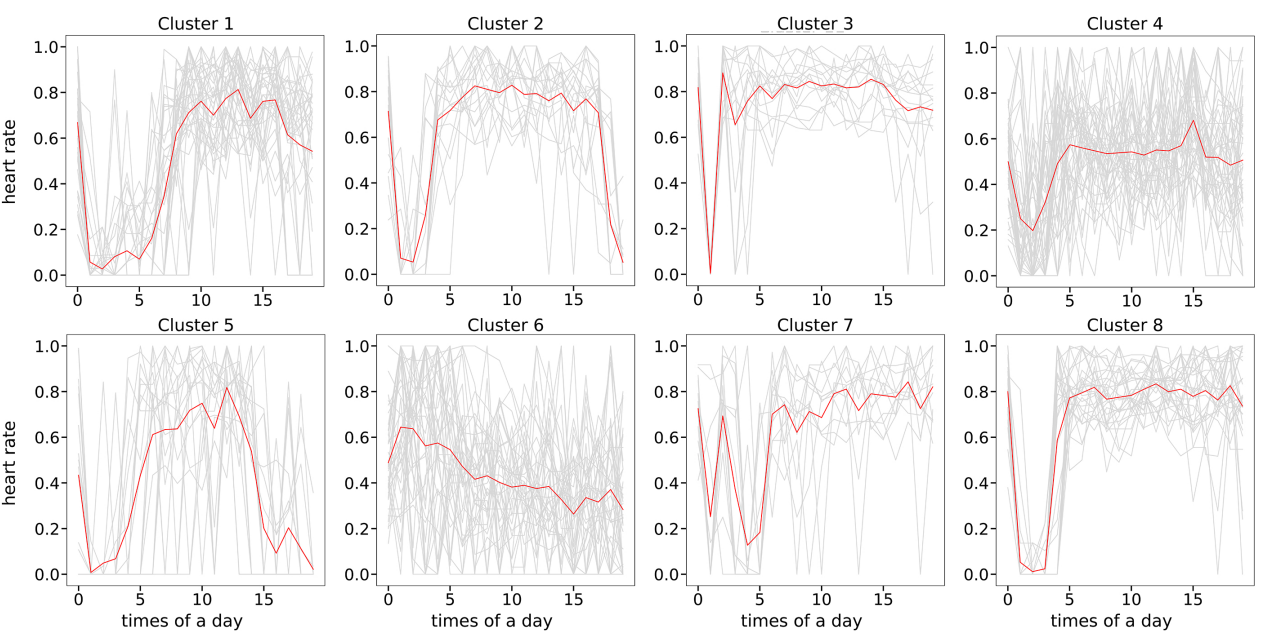}
  \caption{Results of heart rate clustering for weekdays, by using SOM algorithm.}
  \label{fig:wd}
\end{figure}

Moreover, we study the heart rate on weekdays which is shown in Figure \ref{fig:wd}. The valley pattern also exists during this time period, and the peak pattern (cluster 6) i.e. 17\%, happens on weekdays.

\subsubsection{Physical activity Analysis}
To analyze the physical activity of users we use statistical methods. This is due to the fact that clustering algorithms with different parameters and settings do not provide consistent results and patterns. 
Unless specifically noted, all results reported in this section are statistically significant ($p-value < 0.5$).

\begin{figure}[H]
  \centering
  \includegraphics[width= 8 cm ]{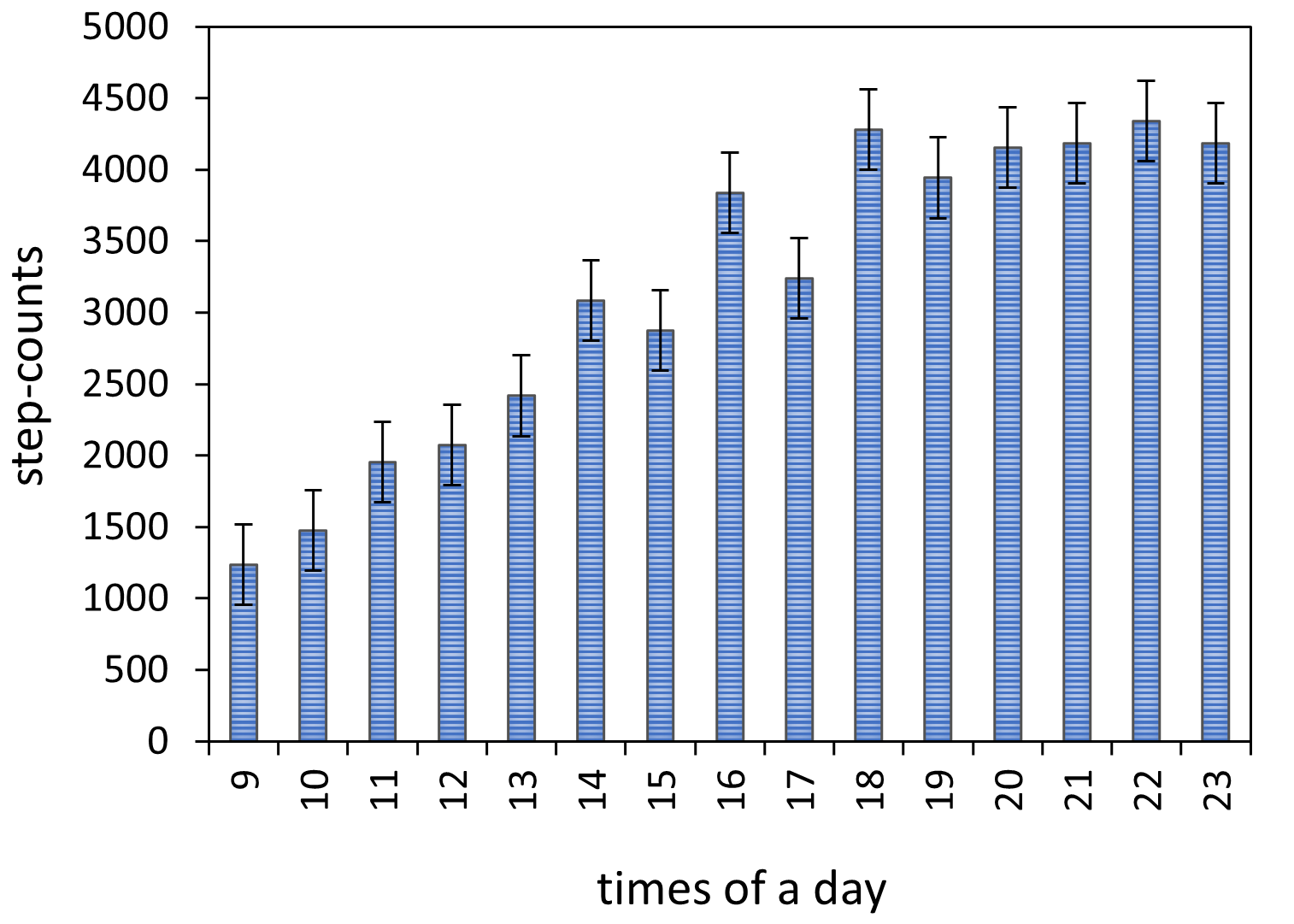}
  \caption{The average amount of physical activity, in terms of steps taken, during a day.} 
  \label{fig:a0}
\end{figure}

\begin{figure}[H]
  \centering
  \includegraphics[width= 8 cm]{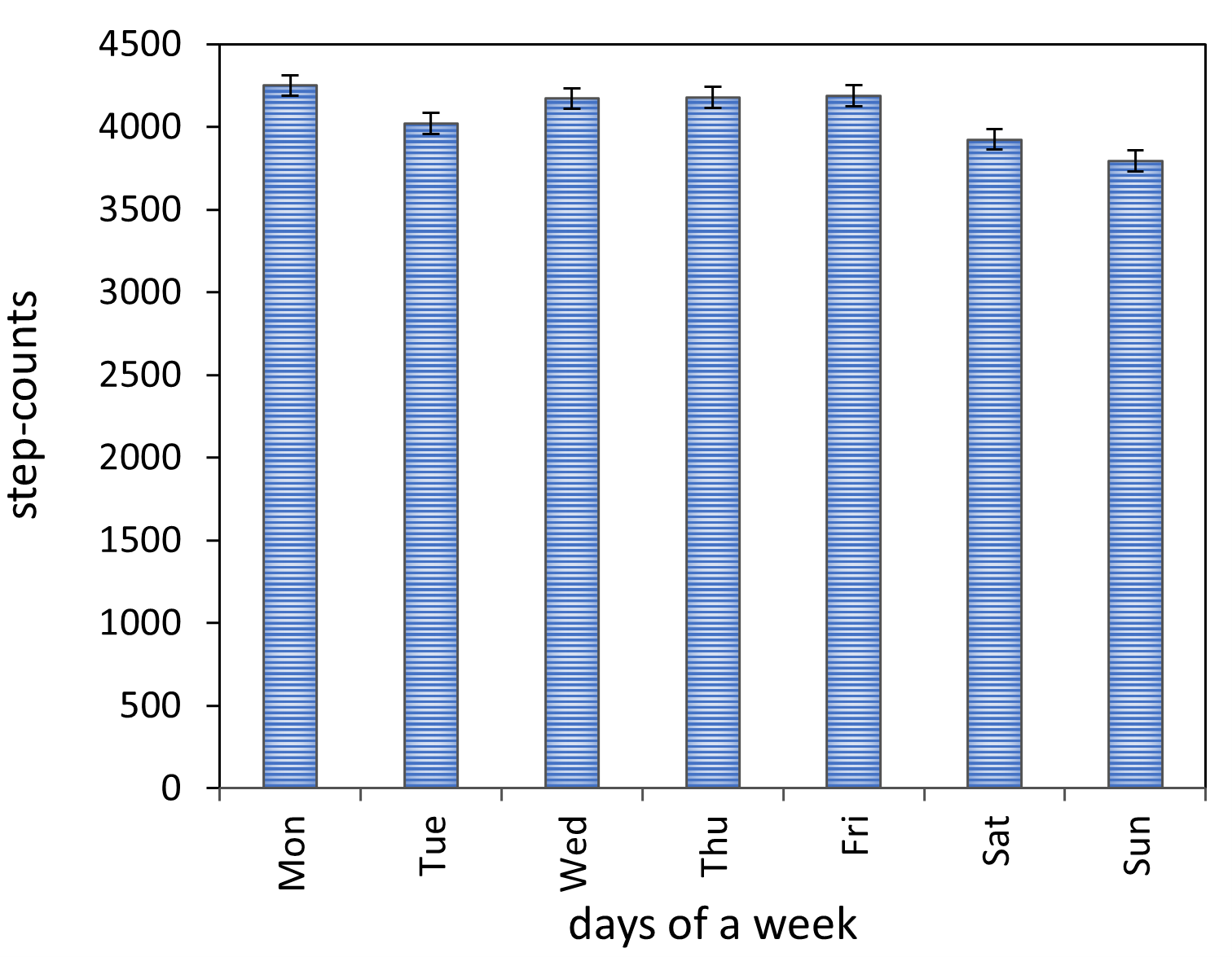}
  \caption{The average number of steps taken, during the day of the week. On weekdays, the number of steps is higher than on weekends.}
  \label{fig:a2}
\end{figure}
\begin{figure}[H]
  \centering
  \includegraphics[width= 8 cm]{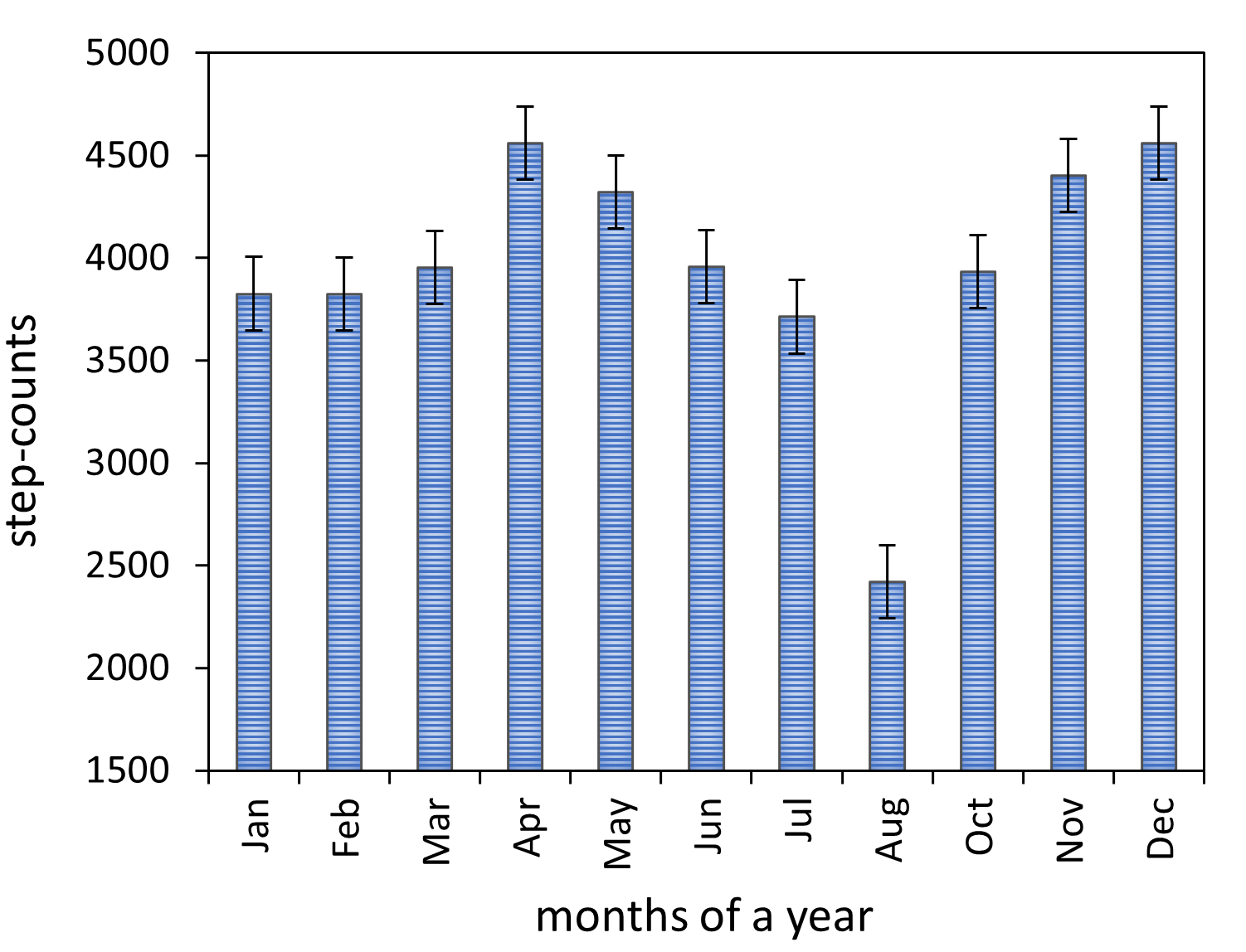}
  \caption{Distribution of physical activities (number of steps taken) by months of a year.}
  \label{fig:a3}
\end{figure}

As shown in Figure\ref{fig:a0}, our statistical analysis reveals that physical activity is increasing during the day, and the highest activity time is at 6:00 p.m. and at the end of the day. We speculate that this pattern is the routine exercise time that users usually perform after working hours.  
We found that users significantly increase physical activities in the afternoon, and after that, step counts in the morning are higher than at night.
Figure \ref{fig:a2} presents the number of weekday physical activities. We can conclude physical activity on the weekend are less than on weekdays.
By investigating physical activity in different months, we found that $p-value$ in August, September, October, and November were not significant. Based on the result presented in Figure \ref{fig:a3} statistical analysis the rest of the months show physical activity is the most at the beginning of the new year.

\subsection{Limitations}
Although we analyzed a large dataset of smartwatch users, our research has several limitations. To fully respect users' privacy, the application does not enforce or motivate them to provide any demographic information. Therefore, our dataset lacks information about users' gender, age, and medical conditions. 

Our dataset can only estimate for the user’s region and is not equipped with a precise geographical location. There, we would not be able to evaluate the climate and community (urban versus rural) and their impact on heart rate and physical activities.

Our existing data and clustering algorithm did not specify a common pattern of physical activity among the group who participated in our study. In our analysis, if statistical significance has not been reached we did not show any trend.   

\section{Conclusion}
We analyzed the SW dataset of 1,014 users. Our dataset was collected in the real world from heterogeneous devices; thus the accuracy of the data can not be compared with Holter devices. Excitingly, however, the heart rate variability patterns we identified strongly resemble Holter devices. We identify several temporal patterns in heart rate variability and physical activities using clustering and statistical analysis. We believe our findings can be desirable for general health practitioners, application designers, device producers, and smartwatch users.
In the future, we intend to focus on clinical settings and identify temporal dynamics of heart-rate variability and physical activity associated with disease and therapeutic modalities such as pharmacotherapy, radiotherapy, and others.

\bibliographystyle{unsrt}  
\bibliography{main}

\end{document}